\documentclass{ws-procs961x669}
\usepackage{graphicx}
\usepackage{chapterbib}
\def\be{\begin{equation}}
\def\ee{\end{equation}}
\def\bea{\begin{eqnarray}}
\def\eea{\end{eqnarray}}
\begin{document}
\title{Time is entropy: A geometric proof}

\author{Hernando Quevedo}

\address{Instituto de Ciencias Nucleares, Universidad Nacional Aut\'onoma de M\'exico,  Mexico City.\\
Dipartimento di Fisica and Icra, Universit\`a di Roma “La Sapienza”, Roma, Italy. \\
Al-Farabi Kazakh National University, Al-Farabi av. 71, 050040 Almaty, Kazakhstan.}

\date{\today}

\begin{abstract}
We analyze the equilibrium space of an ideal gas using the formalism of geometrothermodynamics. We introduce the concept of thermodynamic geodesics to show that the equilibrium space around a particular initial state can be divided into two regions, one that can be reached using thermodynamic geodesics and the second one forbidden by the second law of thermodynamics. Moreover, we show that, along thermodynamic geodesics, entropy is a linear function of the affine parameter, indicating that it can be used as a time parameter with a particular arrow of time determined by the direction in which entropy increases. We argue that entropy can also be interpreted locally as time in the case of any thermodynamic system in equilibrium and systems described within the scope of linear non-equilibrium thermodynamics.  

\end{abstract}


\bodymatter

\section{Introduction}
\label{sec:int}

The relationship between time and entropy has been a subject of intense research since entropy was introduced in classical thermodynamics in the XIX century. The second law of thermodynamics establishes that the entropy of an isolated system increases or remains constant, the last case corresponding to a state of maximum disorder. So far, all the experimental evidence confirms the validity of the increase of entropy as long as the conditions for fulfilling the second law are satisfied. Thus, the second law establishes a particular direction in which the entropy of an isolated thermodynamic system increases and forbids any processes that could lead to a decrease in entropy.

On the other hand, the notion of time plays an essential role in science as a parameter that determines the direction in which all physical processes occur. This particular direction establishes the arrow of time. Moreover,  no experimental evidence has been found so far that contradicts the existence of the arrow of time. 

From the arguments presented above it seems obvious to ask the question about the relation between the direction in which the entropy increases and the arrow of time. Many works have been dedicated to studying this question, which, due to the lack of concrete experimental results, has become an active field of research in philosophy 
\cite{bardon2024brief,grandy2008entropy,lebowitz1993boltzmann,maccone2009quantum,amin2012arrow}.

In this work, we propose to use the approach of geometrothermodynamics (GTD)
\cite{quevedo2007geometrothermodynamics} to relate time and entropy. In this approach, the concepts of classical equilibrium thermodynamics are expressed in terms of concepts of differential geometry in an invariant way. In GTD, as in thermodynamic geometry \cite{weinhold2009classical,ruppeiner1995riemannian}, the equilibrium states of a thermodynamic system are interpreted as points of an abstract space called equilibrium space, whose geometric properties should offer information about the physical properties of the system. The equilibrium space should respect the symmetry properties of classical thermodynamics, namely, it should be independent of the choice of representation and thermodynamic potential \cite{callen1998thermodynamics}. To fulfill this condition, GTD considers the equilibrium space as a subspace of the phase space, in which the changes of representation and thermodynamic potential are represented as coordinate transformations that leave the geometric properties of the phase space invariant.

We investigate the properties of the equilibrium space of the ideal gas, which turns out to be flat. This result allows us to find a simple form for the corresponding geodesic trajectories and show that entropy is equivalent to time if that the laws of thermodynamics are satisfied along the geodesics.

This work is organized as follows. In Sec. \ref{sec:gtd}, we review the main ideas of GTD and show how to incorporate Legendre invariance into the geometric formalism. In Sec. \ref{sec:ig}, we present the explicit form of the metrics for the equilibrium space in the case of a system with two thermodynamic degrees of freedom. We apply this formalism to the case of an ideal gas and show that the corresponding space of equilibrium states is flat. In Sec. \ref{sec:geo}, we analyze the geodesics of the ideal gas, introduce the concept of thermodynamic geodesics, and show that the equilibrium space of the ideal gas is characterized by a geometric structure that resembles the causal structure of special relativity. In Sec. \ref{sec:timent}, we show that entropy can be used as an affine parameter along thermodynamic geodesics, leading to a geometric equivalence between entropy and time. Finally, in Sec. \ref{sec:con}, we present a summary of our results. 
\section{Geometrothermodynamics}
\label{sec:gtd}

Consider a system with $n$ thermodynamic degrees of freedom. To describe such a system in classical thermodynamics, one needs $n$ extensive variables $E^a$ $(a,b,...=1,2,...,n)$, $n$ intensive variables $I_a$ and a thermodynamic potential $\Phi$ \cite{callen1998thermodynamics}, which is usually the entropy or the internal energy.

To describe the properties of the system, we can use two equivalent approaches. The first one uses the equations of state, which relate the intensive and extensive variables $I_a=I_a(E^b)$ and are usually derived by using empirical methods. 
The second approach is based upon the use of the fundamental equation, which relates the thermodynamic potential with the extensive variables $\Phi=\Phi(E^a)$. In addition, the function determining the fundamental equation is assumed to satisfy the first law of thermodynamics 
\be
d\Phi = \sum_a I_a d E^a,
\ee
from which the equations of state can be obtained as
\be
I_a = \frac{\partial \Phi}{\partial E^a} \ .
\ee
In this work, we will use the approach based on the fundamental equation to describe thermodynamic systems.

In equilibrium thermodynamics, there are two preferred thermodynamic potentials, namely, the entropy $S$ and the internal energy $U$ of the system, which are the basis of the so-called entropic and energetic representations, respectively \cite{chur90}. Both representations are equivalent in the sense that the physical properties of a system do not depend on the choice of representation. We consider this property as a symmetry of classical thermodynamics. Furthermore, in classical thermodynamics, we are allowed to use any thermodynamic potential that is obtained from the entropy or the internal energy using  Legendre transformations, without affecting the physical properties of the system. In GTD, we also consider the invariance with respect to Legendre transformations as an essential symmetry of thermodynamics. 

One of the objectives of GTD is to construct a geometric formalism that takes into account the symmetries of classical thermodynamics and represents physical properties in terms of geometric concepts. To this end, we consider an equilibrium state defined as the set of values of $E^a$ that satisfy the fundamental equation $\Phi=\Phi(E^a)$. Furthermore, we represent each equilibrium state as a point of an abstract space that we call equilibrium space ${\cal E}$. This means that ${\cal E}$ is an $n-$dimensional space that can be coordinatized by the set of variables $E^a$. It is then possible to show that ${\cal E}$ can be endowed with a differential structure, which is then used to introduce the corresponding tangent and cotangent spaces.

As mentioned above, in classical thermodynamics, all the thermodynamic properties of a system can be derived from the corresponding fundamental equation  $\Phi = \Phi(E^a)$. We follow this prescription and demand that all the properties of the equilibrium space should be derived from the fundamental equation. On the other hand, a Legendre transformation in ${\cal E}$ consists in introducing a new potential $\tilde \Phi$ from $\Phi$ by means of the relationship
\be
\tilde \Phi = \Phi - \sum_i \frac{\partial \Phi}{\partial E^i} E^i ,
\label{flaw}
\ee
where $i=\{1,...,j\}$ with $j\leq n$. If $j<n$, the transformation is called partial, and for $i=n$ we obtain a total Legendre transformation. 

The above expression shows that it is not possible to interpret a Legendre transformation on ${\cal E}$ as a coordinate transformation because it involves the function $\Phi(E^a)$ and its derivatives. This problem can be solved by introducing the auxiliary $(2n+1)-$dimensional phase space ${\cal T}$ with coordinates $Z^A=\{\Phi, E^a, I_a\}$ $(A=0,1,...,2n)$ so that all the coordinates $Z^A$ are independent from each other
\cite{hermann1973geometry}. Then, a Legendre transformation can be represented as a coordinate transformation by means of the following relationships
\cite{arnolʹd2006mathematical}
\be
\{Z^A\}\longrightarrow \{\widetilde{Z}^A\}=\{\tilde \Phi, \tilde E ^a, \tilde I ^ a\}\ ,
\ee
\be
 \Phi = \tilde \Phi - \sum_i I_i E^i \ ,\quad
 E^i = - \tilde I ^ {i}, \ \  
E^j = \tilde E ^j,\quad   
 I^{i} = \tilde E ^ i , \ \
 I^j = \tilde I ^j \ ,
 \label{leg}
\ee
where $i\cup j$ is any disjoint decomposition of the set of indices $\{1,...,n\}$,
and $k,l= 1,...,i$. In particular, for $i=\{1,...,n\}$ and $i=\emptyset$, we obtain
the total Legendre transformation and the identity, respectively. 

One important property of introducing the phase space ${\cal T}$ is that it is canonically equipped with a contact structure defined as follows. Let $T({\cal T})$ denote the tangent space of ${\cal T}$ and let  ${\cal V}\subset T({\cal T})$ be an arbitrary field of hyperplanes on ${\cal T}$. Then,  there exists a non-vanishing differential 1-form $\Theta$ on the cotangent manifold 
$T^*({\cal T})$ such that the field ${\cal V}$ can be identified with the kernel of $\Theta$, i. e.,
${\cal V} = \ker \Theta$. Furthermore, if the Frobenius integrability condition 
\be
\Theta\wedge d \Theta =0
\ee
is satisfied, the hyperplane field ${\cal V}$ is said to be 
completely integrable. On the contrary, if the condition   
\be
\Theta \wedge d \Theta \neq 0 
\ee
is satisfied, then ${\cal V}$ is called non-integrable. 
The limiting case 
\be
\Theta \wedge (d \Theta)^n \neq 0
\ee
corresponds to a  hyperplane field ${\cal V}$ that is  maximally non-integrable
and is said to define a contact structure on ${\cal T}$. 
The pair $({\cal T},\Theta)$ determines a contact manifold \cite{dillen1999handbook}. Using the coordinates $Z^A$ defined above, according to Darboux theorem \cite{dillen1999handbook}, the contact 1-form can be expressed as
\be
\Theta = d\Phi-\sum_a I_a E^a .
\ee
This canonical representation is particularly important because, under the action of a Legendre transformation of the form (\ref{leg}), it transforms as 
\be
\Theta\rightarrow \tilde \Theta = d\tilde \Phi - \sum_a \tilde I_a \tilde E^a,
\ee
i. e., it is Legendre invariant in the sense that its functional dependence on the coordinates of the phase space does not change under the action of Legendre transformations.  

Finally, we endow the phase space ${\cal T}$ with a Riemannian metric structure $G$, which we demand to be Legendre invariant. The triad $({\cal T}, \Theta, G)$ constitutes a Legendre invariant Riemannian contact manifold. It is the main geometric ingredient of GTD because it contains all the information regarding the invariance with respect to Legendre transformations and changes of representation.

The equilibrium space ${\cal E}$  in GTD is constructed from the triad $({\cal T},\Theta, G)$ as follows. Let ${\cal E}$ be a submanifold of ${\cal T}$ defined by the embedding map 
\be
\varphi: {\cal E}\to {\cal T}
\ee
or in coordinates
\be 
\varphi: \{E^a\} \mapsto Z^A(E^b) = \{\Phi(E^b), E^b, I_a(E^b)\} .
\label{coord}
\ee 
In addition, we demand that  the pullback of 
$\varphi$ annihilates the contact 1-form $\Theta$, i. e., 
\be 
\varphi^*(\Theta)  = \varphi^*(d\Phi - \sum_a I_a E^a)=0.
\ee
This condition implies that on the equilibrium space the first law of thermodynamics (\ref{flaw}) is identically satisfied. Notice that the embedding map $\varphi$, as represented in Eq.(\ref{coord}), implies that the fundamental equation $\Phi = \Phi (E^a)$ must be given explicitly in order for the equilibrium manifold to be well defined. Finally, one can prove that the pullback
 $\varphi^*$ induces a Riemannian metric $g$ on ${\cal E}$ by means of $g=\varphi^*(G)$ or, equivalently,
\be 
g=g_{ab} dE^a dE^b \ , \qquad g_{ab}= \frac{\partial Z^A}{\partial E^a} \frac{\partial Z^B} {\partial E^b} G_{AB} \ .
\label{gdown}
\ee
It then follows that the equilibrium space is a Riemannian manifold $({\cal E},g)$ whose properties are completely determined by the properties of the phase space $({\cal T},\Theta,G)$ and the embedding map $\varphi: {\cal E} \to {\cal T}$. 

The above description of the geometric structure of GTD shows that the phase space metric 
$G_{AB}$, which should be Legendre invariant, and the embedding map $\varphi$, which determines the fundamental equation $\Phi = \Phi(E^a)$, are the starting point to perform the explicit analysis of the corresponding thermodynamic system. To find Legendre invariant metrics in ${\cal E}$, it is necessary to solve a system of algebraic equations for the components of the metric $G_{AB}$ \cite{quevedo2007geometrothermodynamics}. These algebraic equations follow from the condition that the functional dependence of the metric components does not change under the action of Legendre transformations.
 It turns out that  there exist three different classes of Legendre invariant metrics whose line elements can be expressed as follows \cite{quevedo2023unified}
\be
G^{^{I}}=  (d\Phi - I_a d E^a)^2 + (\beta_{ab} E^a I^b) (\delta_{cd} dE^c dI^d) \ ,
\label{GI}
\ee
\be 
G^{^{II}}= (d\Phi - I_a E^a)^2 + (\beta_{ab} E^a I^b) (\eta_{cd} dE^c dI^d) \ ,
\label{GII}
\ee
\be	
\label{GIII}
G^{{III}}  =(d\Phi -  I_a d E^a)^2  + \sum_{a=1}^n \beta_a  E^a I^a    d E^a   d I^a \ ,
\ee
where for simplicity we define $I^a=\delta^{ab}I_b$ and assume the convention of summation over repeated indices. Moreover,  $\delta _{ab}= {\rm diag}(1,1,\cdots,1)$, $\eta_{ab}= {\rm diag}(-1,1,\cdots,1)$, $\beta_{ab}={\rm diag}(\beta_1,\cdots, \beta_n)$. Here,    $\beta_a$ 
are the quasi-homogeneity coefficients of the fundamental equation, i.e., the constants that satisfy the condition  
\be
\Phi(\lambda^{\beta_a} E^a) = \lambda^ {\beta_\Phi}\Phi(E^a)
\label{quasi}
\ee
for real values of $\lambda$ and $\beta_a$. The particular case $\beta_a=1$ and $\beta_\Phi=1$ corresponds to homogeneous functions of degree one, which are commonly used in classical thermodynamics to describe ordinary thermodynamic systems. The coefficients $\beta_a$ determine the Euler identity in the following form \cite{quevedo2019quasi}
\be
\sum_{a=1}^n \beta_a I^a E^ a = \beta_\Phi \Phi \ ,
\label{eul}
\ee
which is used in GTD to investigate the phase transition structure of thermodynamic systems.

As mentioned above, the embedding map $\varphi$ practically determines the thermodynamic system to be considered. In the next section, we will investigate a particular example of this procedure.

\section{Two-dimensional geometrothermodynamics}
\label{sec:ig}

Consider a system with two thermodynamic degrees of freedom ($n=2$) in the entropic representation, i.e., $\Phi=S$, $E^1=U$, and $E^2=V$. The corresponding phase space ${\cal T}$ is 5-dimensional and can be described by the set of coordinates $Z^A=\{S,U,V,I_U,I_V\}$. It is then easy to write down the explicit form of the line elements (\ref{GI}), (\ref{GII}), and (\ref{GIII}), which determine all the geometric properties of the phase space ${\cal T}$ and the equilibrium subspace ${\cal E}$. In fact, as described in the previous section, the equilibrium subspace is completely determined by the embedding map $\varphi: {\cal E}\to {\cal T}$, which in coordinates implies that 
\be 
\varphi: \{U,V\} \mapsto \{S(U,V), U, V, I_U(U,V),I_V(U,V)\}\ .
\label{varphi}
\ee
On the other hand, the condition $\varphi^*(\Theta) =0$ leads to
\be
\varphi^*(\Theta) = \varphi^*(dS - I_U dU - I_V dV) = 0
\ee
so that on ${\cal E}$ it holds that 
\be
dS= I_U dU + I_Vd V\ , \quad I_U = \frac{\partial S}{\partial U}, \quad 
I_V = \frac{\partial S}{\partial V},
\ee
which correspond to the first law of thermodynamics and the equations of state  for this particular case. 

The embedding (\ref{varphi}) implies that the corresponding thermodynamic system is described by the fundamental equation $S=S(U,V)$. As a particular case, we assume that $S(U,V)$ is a homogeneous function so that the coefficients of homogeneity that appear in the line elements $G^I$, $G^{II}$, and $G^{III}$ are simply $\beta_S=\beta_U=\beta_V=1$. It is then straightforward to calculate explicitly the metric components $g_{ab}$ by using Eq.(\ref{gdown}). We obtain obtain
\be
\label{first} 
g^{I}= S \left(
\frac{\partial^2  S }{\partial  U ^2}
d U ^2+2\frac{\partial^2  S }{\partial  U 
	\partial  V  } d U  d V +\frac{\partial^2
	S }{\partial  V ^2} d V ^2\right)\,,
\ee

\be
\label{second} g^{II}= S \left(-
\frac{\partial^2  S }{\partial  U ^2}
d U ^2+\frac{\partial^2  S }{\partial  V ^2}
d V ^2\right)\, ,
\ee

\be
g^{III}=  U  \frac{\partial S }{\partial  U } \frac{\partial^2 S }{\partial  U ^2} d U  ^2
+
S \frac{\partial^2  S }{\partial U \partial  V } d U  d V 
+  V  \frac{\partial  S }{\partial  V } 
\frac{\partial^2  S }{\partial V ^2} d V  ^2\ .
\label{third}
\ee
This result shows that to carry out the explicit calculation of the metric components, we only need to know the explicit form of the fundamental equation $S=S(U,V)$.

The curvature tensor and the scalar curvature of the equilibrium space can be straightforwardly calculated for the above metrics. From the point of view of GTD, the curvature of ${\cal E}$ contains information about the phase transition structure of the corresponding thermodynamic system. In fact, the curvature singularities should indicate the places where phase transitions can occur. The reason for this equivalence is as follows. During a phase transition, the approach of equilibrium thermodynamics breaks down, and non-equilibrium effects should be taken into account. At the level of the geometric description of thermodynamics, a similar behavior should occur, i.e., the geometric description should break down, which is exactly what occurs at the locations where curvature singularities appear. This means that GTD is not able to handle non-equilibrium states in its present formulation. A generalization of GTD to include linear non-equilibrium thermodynamics is currently under construction and will be presented elsewhere.

By using the properties of the metrics $g^I_{ab}$, $g^{II}_{ab}$, and $g^{III}_{ab}$ and the Euler identity as given in Eq.(\ref{eul}), it can be shown that the curvature singularities are determined by the conditions
\bea 
R^I \rightarrow \infty &  \Longleftrightarrow &
\frac{\partial^2  S }{\partial  U ^2} \frac{\partial^2 S }{\partial  V ^2} - \left(\frac{\partial^2 S }{\partial U \partial V }\right)^2 =  0   ,
 \\
R^{II} \rightarrow \infty &  \Longleftrightarrow &
\frac{\partial^2  S }{\partial  U ^2} = 0 \quad {\rm or} \quad  \frac{\partial^2 S\  }{\partial  V ^2} =0 ,
\\
R^{III} \rightarrow \infty & \Longleftrightarrow &
\frac{\partial^2 S }{\partial U \partial V } =0  \ .
\eea
Notice that in two-dimensional Riemannian manifolds, the curvature tensor possesses only one independent component that determines the behavior of the Ricci scalar, which, in turn, enters the expressions for all curvature invariants. For this reason, we limit ourselves to the analysis of the Ricci scalar.  

One can see that the singularities of $R^I$ correspond to the places where the equilibrium condition for a system with two thermodynamic degrees of freedom is violated, which is usually associated with the presence of first-order phase transitions  \cite{callen1998thermodynamics,huang2008statistical}. Furthermore, the singularities of $R^{II}$ and $R^{III}$ are determined by the zeros of the second derivatives of the fundamental equation $S=S(U,V)$, which in classical thermodynamics are associated with divergences of the response functions of the system, implying the presence of second-order phase transitions \cite{callen1998thermodynamics,huang2008statistical,huang2009introduction}.

To exemplify the metric structure of the equilibrium space, let us consider the simple example of an ideal gas for which the fundamental equation can be written as
\be
		S(U,V) = k_B N \left( c_V \ln \frac{U}{NU_0} + \ln \frac{V}{NV_0}\right)\ ,
		\ee
where $k_B$ is the Boltzmann constant $N$ is the number of particles of the system, $c_V$ is the heat capacity, and $U_0$ and $V_0$ are constants. We assume that the number of particles is constant so that the corresponding equilibrium spaces is two-dimensional.

The computation of the metrics of the equilibrium space leads to compatible results. For concreteness, we consider only the line element $g^{III}$. We obtain  
\be
				-g^{III}=ds_{ig}^2= k_B^2 N^2\left( c_V^2 \frac{dU^2}{U^2} + \frac{dV^2}{V^2}\right) .
\label{metig}
\ee
It is then easy to show that the Riemann curvature tensor vanishes, $R_{abcd}^{III}=0$, indicating that the equilibrium space is flat, which is interpreted in GTD as due to the lack 
of thermodynamic interaction. This result agrees with the physical properties of an ideal gas, for which it is well known that no thermodynamic interaction exists between the particles of the gas.

\section{Thermodynamic geodesics}
\label{sec:geo}

Since the equilibrium space of the ideal gas is flat, it is possible to introduce Cartesian-like coordinates in which the thermodynamic metric  (\ref{metig}) takes an Euclidean form. In fact, the coordinate transformation 
\be
\xi = \xi_{int} + k_B N c_V \ln U \ , \quad \eta = \eta_{int}+ k_B N \ln V,
\label{newcoo}
\ee
where $\xi_{int}$ and $\eta_{int}$ are integration constants, transforms the metric (\ref{metig}) into the form
\be
ds_{ig}^2 = d\xi^2 + d\eta ^2 .
\label{flat}
\ee
Notice that the integration constants $\xi_{int}$ and $\eta_{int}$ can be chosen such that 
\be
\xi\geq 0 , \quad  \eta\geq 0
\ee
for all valid values of $U$ and $V$.  This is a very important consequence of this coordinate transformation, which, as we will see below,  allows us to find the intrinsic physical structure of the equilibrium space.

\begin{figure}
\includegraphics[scale=1]{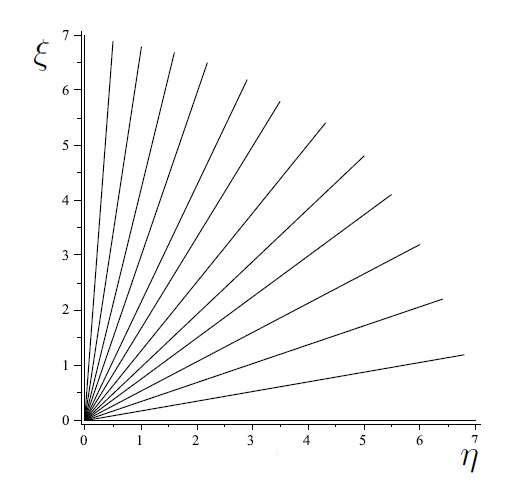}
\label{fig1}
\caption{Geodesics of the equilibrium space of an ideal gas with initial state $\xi_i=0$ and $\eta_i=0$. } 
\end{figure}

Let us define in the equilibrium space ${\cal E}$ the thermodynamic length as \cite{amari2012differential}
\be
L = \int_{p_1}^{p_2} ds_{ig}  dUdV  ,
\ee
where $dU\,dV$ is  the volume element and $p_1$ and $p_2$  represent two arbitrary points in  ${\cal E}$. 
Let us consider only those trajectories between $p_1$ and $p_2$ that satisfy the condition that the thermodynamic length be extremal, $\delta L =0$, i.e., the trajectories correspond to geodesics on ${\cal E}$. In the case of the flat metric (\ref{flat}), 
the geodesic equations become
\be 
\frac{d ^2 \xi}{d\tau^2} =0\ , \quad \frac{d^2 \eta}{d\tau^2} = 0,
\ee
where $\tau$ is the affine parameter along the geodesics. The solution of these geodesic equations can be written as
\be
\xi = \xi_0 + \xi_1 \tau\ , \quad \eta = \eta_0 + \eta_1 \tau \ ,
\ee 
where $\xi_0$, $\xi_1$, $\eta_0$, and $\eta_1$ are real constants, satisfying the condition $\xi\geq 0 $ and $\eta \geq 0$. The linearity of $\xi$ and $\eta$ as functions of $\tau$ can be used to represent the geodesics as straight lines, $\xi =c_0 + c_1 \eta$, on the plane $(\xi,\eta)$, where $c_0$ and $c_1$ are constants. If we consider,  for instance, all the geodesics with initial state $\xi_i=0$ and $\eta_i=0$, then they must be contained within the region determined by the allowed values  $\xi\geq 0 $ and $\eta \geq 0$, as shown in Fig. \ref{fig1}. In fact, for any point $(\xi_f,\eta_f)$ of ${\cal E}$,  one can always find a geodesic that connects it with the origin ($\xi_i=0,\eta_i=0)$. This means that the entire equilibrium space can be covered with geodesics that start from the point $(0,0)$. 

We now consider the question of the physical significance of the geodesics. In classical thermodynamics, two equilibrium states can be connected with each other by quasi-static processes, i.e., a sequence of equilibrium states. We use this criterion to introduce the notion of thermodynamic geodesics, i.e., geodesics whose points can be connected by quasi-static processes and along which the laws of thermodynamics are fulfilled. Since the thermodynamic system is determined by the fundamental equation, the first law is identically satisfied. The second law implies that only those geodesics are allowed along which the entropy increases. The third law indicates that the point of the equilibrium space that corresponds to the state of minimum entropy must be excluded from consideration. This is an interesting result since it indicates that the equilibrium space is not trivial from a topological point of view. Accordingly, the origin of coordinates plays an important role in determining the local and global properties of the equilibrium space.

\begin{figure}
    \centering
\includegraphics[scale=1.2]{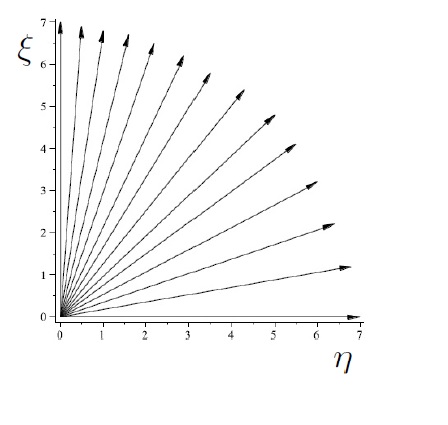}
    \caption{Thermodynamic geodesics with a preferred direction as demanded by the second law of thermodynamics.}
    \label{fig2}
\end{figure}

\begin{figure}
    \centering
\includegraphics[scale=1.1]{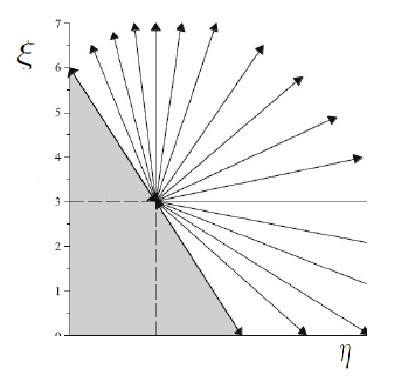}
    \caption{Thermodynamic geodesics with an initial state at $(\xi_i=3,\eta_i=2)$. States in the gray zone cannot be reached by thermodynamic geodesics.}
    \label{fig3}
\end{figure}

We will now consider the consequences of the second law. Thermodynamic geodesics must have a definite direction determined by the increase of entropy. This is illustrated in Fig. \ref{fig2}. 
Again, we can see that it is possible to cover the entire equilibrium space by thermodynamic geodesics with the initial state ($0,0$).
The situation is different when we consider a non-zero initial state, as illustrated in Fig. \ref{fig3}.
Geodesics connecting the initial state $(\xi_i=3,\eta_i=2)$ with any state inside the gray zone imply a decrease of entropy, i.e., they cannot be considered thermodynamic geodesics. This structure is valid for any initial state different from the zero state, which is, however, forbidden by the third law of thermodynamics because it corresponds to a state of minimum entropy. It follows that, in general for any initial state, the equilibrium space is divided into two regions, one which contains thermodynamic geodesics representing quasi-static processes, and a second region, which is non-physical in the sense that no thermodynamic geodesic can reach any state within this region. This split of the equilibrium space resembles the causal structure of special relativity \cite{quevedo2015relativistic}. To show this,  we first notice that  
 it is possible to use the entropy as an affine parameter as follows. Introducing the geodesic solutions 
$\xi=\xi_0 + \xi_1 \tau$ and  $\eta=\eta_0 + \eta_1 \tau$ into the fundamental equation of the ideal gas, we obtain $S=\tilde s _0 + \tilde s_1 \tau$ so that the affine 
parameter can be represented as $\tau = \tau_0 + \tau_1 S$, where $\tau_0$ and $\tau_1$ are constants. This implies that in terms of $S$ the logarithmic coordinates
can be expressed as
\be
\xi=\tilde \xi_0 + \tilde \xi_1 S\ , \quad \eta=\tilde \eta_0 + \tilde \eta_1 S \ ,
\ee
where
\be
\tilde\xi_0 = \frac{\xi_0\eta_1 - \xi_1 \eta_0 -  s_0  \xi_1}{ \xi_1 + \eta_1 }\ , \quad 
\tilde\xi_1 = \frac{ \xi_1}{ \xi_1 + \eta_1}\ ,
\ee
\be
\tilde\eta_0 = \frac{ \xi_1\eta_0 - \eta_1\xi_0 - s_0  \eta_1  }{ \xi_1 + \eta_1}\ , \quad
\tilde\eta_1 = \frac{\eta_1}{ \xi_1 + \eta_1}\ .
\ee

\begin{figure}
    \centering
    \includegraphics[scale=1.2]{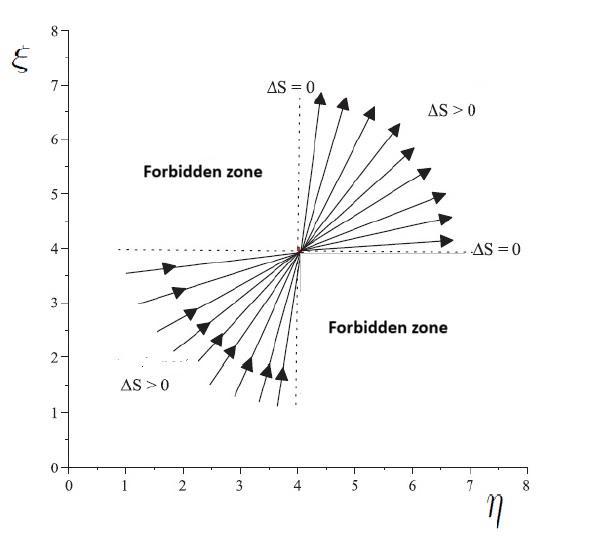}
    \caption{Distribution of thermodynamic geodesics in the equilibrium space.}
    \label{fig4}
\end{figure}

The original constants $\xi_0$, $\xi_1$, $\eta_0$, and $\eta_1$ can be chosen such that the new constants $\tilde \xi_0$, $\tilde \xi_1$, $\tilde \eta_0$, and $\tilde \eta_1$ are  positive definite. 
Then, an arbitrary initial state  $(\xi_i,\eta_i)$ can be connected with  a final state  $(\xi_f,\eta_f)$ by a thermodynamic geodesic,
if the following conditions are satisfied: 
\be
\Delta \xi = \xi_f -\xi_i =\tilde \xi_1 \Delta S \geq 0\ , \quad
\Delta \eta = \eta_f -\eta_i =\tilde \eta_1 \Delta S \geq 0\ , 
\ee
From the second law, we know that  $\Delta S \geq 0$. Then, we conclude that all the thermodynamic geodesics must satisfy the conditions
\be
\Delta\xi\geq 0 , \qquad 
\Delta\eta\geq 0,
\ee
simultaneously. Consequently, all the thermodynamic geodesics that initiate at a particular equilibrium state must be contained within the region defined by $\Delta\xi =0$ and $\Delta\eta =0$.  On the other hand, all the thermodynamic geodesics that end at a particular equilibrium state, say $(\xi_f,\eta_f)$ 
must obey the same conditions. Figure \ref{fig4} illustrates this behavior. We conclude that all the incoming and outgoing thermodynamic geodesics at a given state must be contained within a cone, which is called adiabatic cone  \cite{quevedo2015relativistic}.

\section{Time and entropy}
\label{sec:timent}

The fundamental equation of the ideal gas in the new coordinates $\xi$ and $\eta$ can be expressed as
\be
S = S_0 + \xi + \eta ,
\ee
or in terms of the affine parameter $\tau$ as
\be
S= S_0 +\xi_0 + \eta_0 + (\xi_1+\eta_1) \tau .
\ee
Since an affine parameter is defined modulo a linear transformation, it follows that entropy can be used as an affine parameter along the thermodynamic geodesics. On the other hand, as we argued above, a thermodynamic geodesic can be understood as describing a sequence of equilibrium states, which, in turn, is interpreted in classical thermodynamics as a quasi-static process. It follows that entropy can be used to identify the states of a quasi-static process in such a way that a preferred direction is determined by the second law of thermodynamics, namely, the direction in which the entropy increases, i.e., the ``arrow of entropy" or the ``arrow of the affine parameter". It is easy to show that the affine parameter can, in fact, be used as a time parameter to identify the states of a thermodynamic geodesic, i.e., of a quasi-static process. This shows the equivalence of entropy and time from the point of view of processes that occur within the limits of the applicability of classical thermodynamics. As a consequence, we have established that the arrow of time is a consequence of the second law of thermodynamics.

The above analysis is based on the fact that the equilibrium space of an ideal gas is flat and, consequently, the solutions to the geodesic equations can be represented as straight lines. However, this result can be extrapolated to include from the point of view of GTD any thermodynamic system. In fact, the equilibrium space is a Riemannian differential manifold and as such is locally flat in the sense that any point and its neighborhood can be identified as a subspace of the flat manifold $R^n$. Then, the equivalence between entropy and time, as shown above, can be generalized to include locally the case of any thermodynamic system that can be described by means of a fundamental equation.

The case of non-equilibrium thermodynamics can also be included as a case in which the equivalence between time and entropy is locally valid. Unfortunately, there is no definite theory of non-equilibrium thermodynamics \cite{de2013non,garcia1991extended}. The only case in which a definite formalism has been developed is for systems out of equilibrium, where the  Onsager reciprocal relations are valid and a notion of local equilibrium exists. This can be understood as a linear perturbation of equilibrium thermodynamics, in which non-equilibrium variables are taken into account up to the first order only. This approximation implies that locally all the states can be considered as equilibrium states. It then follows that classical equilibrium thermodynamics is valid locally. This implies that the equivalence between time and entropy described above is valid locally in Onsager's non-equilibrium thermodynamics.

\section{Conclusions}
\label{sec:con}

In this work, we have used the formalism of GTD to show that the equilibrium space of the ideal gas is a two-dimensional Riemannian manifold with a flat metric. We started from the fundamental equation of the ideal gas and derived the corresponding metric, for which we introduced Cartesian-like coordinates that bring it into an Euclidean form. In these coordinates,  the geodesics can be represented as straight lines. 

We also introduced the concept of thermodynamic geodesics, which are geodesic paths along which the laws of thermodynamics are fulfilled. This allows us to interpret quasi-static processes as thermodynamic geodesics. Moreover, we show that the second law endows the equilibrium space with a geometric structure that resembles the causal structure of special relativity.

On the other hand, in this representation, the entropy of the ideal gas turns out to be related to the affine parameter along the thermodynamic geodesics using a linear transformation. In turn, the affine parameter can be used to identify the sequence of equilibrium states along quasi-static processes, i.e., as a time parameter that characterizes the evolution of quasi-static processes. We interpret this result as showing the equivalence between time and entropy from the point of view of the geometric structure of the equilibrium space. We also argue that this result can be generalized to include the equilibrium space of any thermodynamic system in equilibrium and the case of linear non-equilibrium thermodynamics. In these cases, the equivalence between time and entropy is valid locally, i.e., on the topological neighborhood of any equilibrium state. 

We conclude that entropy and time are closely related from the point of view of GTD. In fact, entropy is the parameter that can be used to measure the time that passes between the equilibrium states of quasi-static processes.

\section*{Acknowledgments}

This work was supported by UNAM PASPA-DGAPA.


\end{document}